\documentclass[unsortedaddress]{revtex4}

\usepackage{amsfonts}
\usepackage{graphicx}
\usepackage{amstext}
\usepackage{amsthm}
\usepackage{amsopn}
\usepackage{amsmath}
\usepackage{amssymb}
\usepackage{amscd}

\begin{document}

\title{Some properties of deformed $q$-numbers }
\author{Thierry C. Petit Lob\~ao}
\address{Instituto de Matem\'atica, Universidade Federal da Bahia \\
Campus Universit\'ario de Ondina, 40170-110 Salvador--BA, Brazil}
\email{thierry@ufba.br}
\author{Pedro G. S. Cardoso}
\author{Suani T. R. Pinho}
\address{Instituto de F\'{\i}sica, Universidade Federal da Bahia \\
Campus Universit\'ario de Ondina, 40210-340 Salvador--BA, Brazil}
\author{Ernesto P. Borges}
\address{Escola Polit\'ecnica, Universidade Federal da Bahia \\
Rua Prof.\ Aristides Novis 2, 40210-630 Salvador--BA, Brazil}


\begin{abstract}

Nonextensive statistical mechanics has been a source of
investigation in mathematical structures such as deformed algebraic
structures. In this work, we present some consequences of
$q$-operations on the construction of $q$-numbers for all numerical
sets. Based on such a construction, we present a new product that
distributes over the $q$-sum. Finally, we present different patterns
of $q$-Pascal's triangles, based on $q$-sum, whose elements are
$q$-numbers.


\end{abstract}

\maketitle



\section{Introduction}

The $q$-operations \cite{epb:q-algebra, wang:q-algebra} that emerges
from nonextensive statistical mechanics  \cite{ct:1988} seems to
provide a natural background for its mathematical formulation. The
definitions of $q$-sum and $q$-product, on the realm of real
numbers,
\begin{equation}
 \label{q-sum}
 x \oplus_q y := x + y + (1-q) xy,
\end{equation}
\begin{equation}
 \label{q-product}
 x \otimes_q y := \left[ x^{1-q} + y^{1-q} -1 \right]_+^{\frac{1}{1-q}},
 \quad x>0,\, y>0,
\end{equation}
where $[p]_+=\max\{p,0\}$, 
 allow some expressions of nonextensive statistical mechanics be written
  with the same formal simplicity of
the extensive ($q=1$) formalism. For instance, the $q$-logarithm
\cite{quimica} of a product, and the $q$-exponential of a sum are
written as
$$ \ln_q xy=\ln_q x \oplus_q \ln_q y,$$
$$e_q^{x+y} = e_q^x \otimes_q e_q^y,$$
with
\begin{equation}
 \label{q-log}
 \ln_q x := \frac{x^{1-q}-1}{1-q}, \quad x>0
\end{equation}
and
\begin{equation}
 \label{q-exp}
 e_q^x := [1+(1-q)x]_+^{\frac{1}{1-q}}.
\end{equation}
The $q$-sum and the $q$-product are associative, commutative,
present neutral element (0 for $q$-sum and 1 for $q$-product) and
opposite and inverse elements under restrictions. A reasonable
question is whether those operations provide a structure of
commutative ring or  even field. Since the $q$-product does not
distribute over the $q$-sum, they do not define those algebraic
structures.

There are instances of other structures that are distributive,
though do not present other properties. For instance, the tropical
algebra \cite{tropical} --- for which the $T$-sum of two extended
real numbers (${\mathbb R} \cup \{-\infty\}$) is the minimum between
them and the $T$-product is the usual sum --- does not have
reciprocal elements in relation to the $T$-sum.

On the other hand, the relevant structure of a near-ring \cite{Clay}
is an example of a non-distributive ring; however, in this case,
distributivity is required in at least one side. It is known long
ago, as pointed out by Green \cite{Green}, that practical examples
of (both sides) of non-distributive algebraic structures are not so
easy to find out. So the $q$-algebraic structure is a good example
of both-side non-distributive structure.

Recently, we have generalized the $q$-algebraic structure into a
biparametrized $(q,q')$-algebraic structure (and, more generally,
into a $n$ parameter algebraic structure) \cite{JMP-biparametrized},
in such a way that the two-parameter operators $(q,q')$-sum,
$(q,q')$-product, and their inverses, present the same properties of
the monoparameterized $q$-algebraic structure.

A remarkable feature of these algebraic structures is that the
distributivity property does not hold. Though this ``non-property''
is very interesting, there are some proposals in the literature
\cite{wang:q-algebra,Kalogeropoulos}  (that will be shown later)
which change somehow the $q$-algebraic structure in order to recover
distributivity . In all of those proposals
\cite{wang:q-algebra,epb:q-algebra,Kalogeropoulos}, the operations
are deformed but the numbers are not deformed. In this work, we
deform the numbers to obtain the $q$-numbers $x_q$ for all numerical
sets based on $q$-sum in such way that
\begin{equation}
\label{qsumnumber}
 x_q \oplus_q y_q = (x + y)_q.
 \end{equation}
 Since the $q$-product is, in a sense that we will discuss later,
 intrinsically non-distributive, in order to obtain the distributive
 structure in a very natural way, we keep the $q$-sum and propose a new product such that
\begin{equation}
\label{diamprodnumber} x_q \Diamond_q y_q = ( x \, y)_q.
\end{equation}

We also set up the $a$-numbers and $k$-numbers based on other
deformed sums presented in \cite{wang:q-algebra,Kalogeropoulos}. We
call the attention to the interesting connection between the
$q$-natural number and the Heine number \cite{Heine}. Other
mathematical objects, whose elements are $q$-numbers, may be
generated by deformed operations; we exemplify some $q$-Pascal's
triangles, derived by $q$-sum, that correspond to different
patterns.

The paper is organized as follows: Sec.\ \ref{sec:qnumericalsets}
introduces the $q$-numerical sets; Sec.\ \ref{sec:newproduct}
proposes a different product $\Diamond_q$;  other mathematical
objects as $q$-Pascal's triangles are addressed in Sec.\
\ref{sec:pascaltriangle}. Finally, in Sec.\ \ref{sec:conclusions} we
draw our concluding remarks.
%


\section{The $q$-numerical sets}

\label{sec:qnumericalsets}

The main idea is to use the classical construction of the numerical
sets \cite{numsets} for which elements are the respective deformed
numbers. We use the notation $\mathbb{N}_q$, $\mathbb{Z}_q$,
$\mathbb{Q}_q$, $\mathbb{R}_q$ for $q$-natural, $q$-integer,
$q$-rational and $q$-real numerical sets respectively.

Consider an induction over an arbitrary generator $g$ (that we
assume different from 0 and $-1/(1-q)$ to avoid trivial structures)
$q$-summed $n$ times:
\begin{eqnarray}
\label{induction} g & &  \nonumber \\
 g \oplus_q g & = & 2g + (1-q) g^2 \nonumber \\
g \oplus_q g \oplus_q g & = & 3g + 3(1-q)g^2 + (1-q)^2 g^3  \\
\vdots  \nonumber \\
\underbrace{g \oplus_q \dots \oplus_q g}_{n\,\mathrm{times}} & = &
\frac{[1+(1-q) g]^n -1]}{1-q}.  \nonumber
\end{eqnarray}

For simplicity of the expressions in this note, we shall choose
$g=1$, and obtain the deformed $q$-natural number summed $n$ times:
\begin{eqnarray}
\label{qnatural}
 n_q & = & \underbrace{1 \oplus_q + \dots + \oplus_q
1}_{n\,\mathrm{times}} \nonumber \\
& = & \frac{(2-q)^n-1}{1-q}  \nonumber \\
& = & \sum_{k=1}^{n}
 \left( \begin{array}{ll}
n \\
 k \\
 \end{array} \right)
 (1-q)^{k-1},
\end{eqnarray}
where $\left( \begin{array}{ll} n \\ k \\ \end{array} \right)$
stands for the binomial coefficients. The $q$-neutral element is the
same as the usual one, $0_q=0$ ($n_q \oplus_q 0_q = n_q$), and also
$1_q=1$. Of course $n_q\to n$ as $q\to 1$.

The dependence of the parameter $q$ provides a plethora of
interesting different structures. For instance, with $q=2$, we have
a structure given by $\{1\}$; with $q=3$, we have a structure
isomorphic to the finite field with two elements $\{0,1\}$. However,
if $q<2$, we have infinite structures whose elements are all real
numbers (if complex numbers are allowed, there is more freedom on
the parameter $q$).

It is not difficult to verify that the set ${\mathbb N}_q=\{n_q, n
\in {\mathbb N}\}$ with the map $\sigma : {\mathbb N}_q \rightarrow
{\mathbb N}_q \, \,, n_q  \mapsto n_q \oplus 1_q$ is a model for the
Peano axioms. Let us show, for example, some elements of the set
${\mathbb N}_q$ for $q=0$:
$$\mathbb{N}_0=\{0,1,3,7,15,31,63,127,255,511,1023, \dots\}.$$

From the set of the deformed $q$-natural number, we may construct,
as in the classical way, by means the (difference) equivalence
relation on ${\mathbb N}_q \times {\mathbb N}_q$, the set of
deformed $q$-integer numbers ${\mathbb Z}_q$. We also draw some
elements of this set for $q=0$:
\begin{eqnarray*}
\mathbb{Z}_0=\{\dots,-\frac{127}{128},-\frac{63}{64},-\frac{31}{32},
-\frac{15}{16},-\frac{7}{8},-\frac{3}{4},-\frac{1}{2}, \\
0,1,3,7,15,31,63,127,255,511,1023, \dots\}.
\end{eqnarray*}
It is interesting to note that, in any case, the $q$-integer are
strictly greater than $-1/(1-q)$.

The $q$-integer numbers were also studied by R. Cardo and A.
Corvolan \cite{next2008} based on the $\odot_q$ operation introduced
in \cite{epb:q-algebra}: $n_q=n \odot 1$, which is defined in the
same way as the notion we introduced in (\ref{induction}) by
induction.

Analogously, we have also constructed, as in the classical case, the
deformed $q$-rational numbers ${\mathbb Q}_q$, by an (ratio)
equivalence relation on ${\mathbb Z}_q \times {\mathbb Z}_q^*$, and
the $q$-real numbers, ${\mathbb R}_q$, by Cauchy sequences. It would
also be possible to construct the $q$-real numbers using the
Dedekind cuts.

We have proved that, following the classical construction of the
$q$-real number, they are given by
\begin{equation}
\label{qreal}
 x_q=\frac{(2-q)^x-1}{1-q}.
 \end{equation}

 The asymptotical behavior of $x_q$ ($x \rightarrow \infty$)
is given by:
\begin{eqnarray}
\lim_{x\to\infty} x_q =
   \left\{
      \begin{array}{cl}
         \infty, & q<1 \\
         x,      & q=1 \\
         \frac{1}{q-1}, & 1<q\le2.
      \end{array}
   \right.
\end{eqnarray}
\begin{figure}[!h]
\begin{center}
\includegraphics*[width=6.0cm]{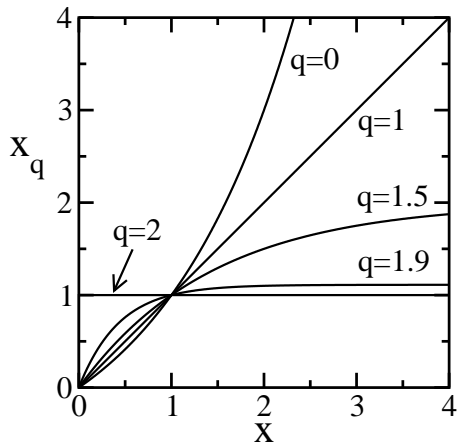}
\end{center}
\caption{\label{fig1} $q$-Real number $x_q$ versus $x$ for some
typical values of $q$.}
\end{figure}
For $q>2$, $x_q$ may assume complex values.

It is amazing to note that in the study of the $q${\it -analogues}
of the hypergeometric series \cite{quantum}, at the second half of
nineteenth century, Heine introduced the deformed number
\cite{Heine}
\begin{equation}
\label{Heine} [n]_H=\frac{H^n-1}{H-1},
\end{equation}
known as the $q${\it -analogue} of $n$. The number deformation plays
a fundamental role in combinatorics, but also have applications in
the study of fractals, hyperbolic geometry, chaotic dynamical
systems, quantum groups, etc. There are also many physical
applications, for instance, in exact models in statistical
mechanics.
It is interesting that the deformed $q$-number (\ref{qreal}) is
exactly the Heine number, by the simple change of variables $q=2-H$.
The connection between nonextensive statistical mechanics and the
Heine number (and quantum groups) was already pointed out in
\cite{Tsallis1994}. It is worth to note that the coincidence of the
symbol $q$ in all these different contexts ($q$-series,
$q$-analogues, $q$uantum groups, and $q$-entropy) occurs just by
chance.

It is possible to define other generalized numbers, based on the
algebraic structures proposed on
\cite{wang:q-algebra,Kalogeropoulos}.
In \cite{wang:q-algebra}, two operations $a$-sum ($+^a$ and $+_a$)
and $a$-product ($\times^a$ and $\times_a$) were introduced. The
$a$-sums are, respectively,
\begin{equation}
\label{niva0-sum}
 x +_a y:=x \oplus_{q} y \,\,\,\, \mbox{with} \,\,\,\, q=1-a,
 \end{equation}
 \begin{equation}
\label{niva-sum}
 x +^{a}y := \left\{ a \ln { \left[ \exp \left( \frac{x^a}{a} \right) +
\exp \left(\frac{y^a}{a} \right) \right]} \right\}^{1/a}.
\end{equation}
The $a$-products are, respectively,
\begin{equation}
\label{niva0-prod}
 x \times_a y := x \otimes_{q} y \,\,\,\, \mbox{with} \,\,\,\, q=1-a,
\end{equation}
\begin{equation}
\label{niva-prod}
 x\times^{a} y:=\frac{\exp{[\ln (1 + a x) \ln(1+a
y)/a}]-1}{a}.
\end{equation}
Based on (\ref{niva-sum}), we obtain the deformed $x^{(a)}$ number
with generator $g$:
\begin{equation}
\label{nniva} x^{(a)}=[a \, \ln x + g^a]^{1/a}.
\end{equation}
 In \cite{Kalogeropoulos}, two operations
$k$-sum ($\boxplus^k$ and $\boxplus_k$) and $k$-product
($\boxtimes^k$ and $\boxtimes_k$) were proposed. The $k$-sums are,
respectively,
\begin{equation}
\label{nikos0-sum}
 x \boxplus_k y := x \otimes_{q} y \,\,\,\, \mbox{with} \,\,\,\, q=1-k,
 \end{equation}
\begin{equation}
\label{nikos-sum}
 x\boxplus^k y := \frac{[(1+kx)^{1/k} + (1+ky)^{1/k}]^k
 -1}{k}.
\end{equation}
The $k$-products, are, respectively
 \begin{equation}
 \label{nikos0-prod}
x \boxtimes_k y := x \oplus_{q=k-1}y  \,\,\,\, \mbox{with} \,\,\,\,
q=1-k,
\end{equation}
\begin{equation}
\label{nikos-prod} x\boxtimes^k y :=
\left[\frac{(xy)^k-x^k-y^k+(k+1)}{k}\right]^{1/k}.
\end{equation}

Based on (\ref{nikos0-sum}) and (\ref{nikos-sum}), the deformed
numbers with generator $g$, $x^{[k]}$  and $x_{[k]}$, associated to
$\boxplus^k$ and $\boxplus_k$, respectively, are:
\begin{equation}
\label{nnikos0} x^{[k]}=\frac{x^k (1 + k \,g) -1}{k},
\end{equation}
\begin{equation}
\label{nnikos0} x_{[k]}=[x \,g^k - (x-1)]^{1/k}.
\end{equation}
%


\section{Distributive property}

\label{sec:newproduct}

The $q$-product is non-distributive, i.e.,
\begin{equation}
 x\otimes_{q}(y+z)\ne
 (x\otimes_{q}y) + (x\otimes_{q}z)\,,
 \forall x\ne 0\,,1\,,\forall q \in \mathbb{R}-\{1\}.
\end{equation}

As an essential result for our work, we observe that, assuming a set
with more than one element and keeping reasonable properties such as
the additive neutral element and cancellation to sum, then there is
no deformed sum that is distributed by the $q$-product. In fact:

Let $t$ be the neutral element of such a sum. If we impose the
distributive property:

$$x \otimes_q (y \oplus t) = (x \otimes_q y) \oplus (x \otimes_q
t)$$
$$ x \otimes_q y = (x \otimes_q y) \oplus (x \otimes_q t)$$
Thus $t = x \otimes_q t$, using (\ref{q-product}), we obtain
$$x^{1-q}=1,$$
i.e., $x$ has to be one of the complex roots $1^{1/(1-q)}$; so,
restricted to real numbers, $x$ has to be 1. Since $x$ is any
element, the set has just one element.

\vspace{0.3cm}

Therefore the non-distributivity is an intrinsic property of the
$q$-product. Some authors \cite{wang:q-algebra,Kalogeropoulos} tried
to obtain distributive structures based on $q$-operations. For
instance, note that, although the operation $\times_a$  is
distributive over $+^a$, shown in (\ref{niva-sum}), $+^a$ does not
have neutral element, as it was consistent with the above result.
Moreover $\times^a$, shown in (\ref{niva-prod}), is distributive
over $+_a$.

 Concerning
the  $k$-sums and the $k$-products, $\boxtimes_k$ is distributive
over $\boxplus^k$, shown in (\ref{nikos-sum}), as well as
$\boxtimes^k$, shown in (\ref{nikos-prod}), is distributive over
$\boxplus^k$. Note that the distributivity results from the curious
exchange of roles of the operations: the $k$-sum $\boxplus_k$ is
indeed a $q$-product, and the $k$-product $\boxtimes_k$ is a
$q$-sum.

 Since there is no deformed sum that is distributed by the
 $q$-product, we propose a new product, signed $\Diamond_q$, that emerges
naturally from the classical construction of the numerical set, just
mentioned. This new product is different from
 equations (\ref{niva-prod}) and (\ref{nikos-prod}), and distributes over the
 $q$-sum. It is defined as
\begin{equation}
\label{diamond-prod}
 x \,\Diamond_q \, y :=
\frac{(2-q)^{\{ \frac{\ln[(1+(1-q) \, x] \ln[1+(1-q) \,y]}{[\ln
(2-q)]^2}\}}-1}{1-q}.
\end{equation}
Moreover the $q$-sum and the $q$-product obey, respectively,
(\ref{qsumnumber}) and (\ref{diamprodnumber}). For the $q$-sum, we
have:
\begin{eqnarray}
\label{qsum1}
 x_q \oplus_q y_q  & = &
\frac{(2-q)^x+(2-q)^y-2}{1-q} \\
 & & + \frac{[(2-q)^x-1][(2-q)^y-1]}{1-q}
 \nonumber \\
 & = &  \frac{(2-q)^{x+y} -1} {1-q}  =  (x+y)_q .
\end{eqnarray}
For the $q$-product, we have
\begin{eqnarray}
\label{qdiamond}
 x_q \Diamond_q y_q  & =
& \frac{(2-q)^{\{ \frac{\ln[(2-q)^x] \ln[(2-q)^y]}{[\ln
(2-q)^2]}\}}-1}{1-q}  \nonumber \\
 & = &  \frac{(2-q)^{x\,y} -1} {1-q}  =  (x \, y)_q.
\end{eqnarray}
It is obvious that, when $q \rightarrow 1$, $x_1 \Diamond_1 y_1 = (
x \, y)_1= x \, y$.

Using (\ref{qsum1}) and (\ref{qdiamond}), it is easy to prove that
the $\Diamond_q$ product distributes over the $q$-sum when applied
to $q$-numbers, i.e.:

\begin{equation}
\label{dist-diamond} x_q \Diamond_q (y_q \oplus_q z_q) = [x_q
\Diamond_q y_q] + [x_q \Diamond_q y_q].
\end{equation}
In other words,
\begin{eqnarray}
x_q \Diamond_q (y_q \oplus_q z_q) & = & x_q \Diamond_q ( y + z)_q  \nonumber \\
\nonumber & = & [x \,(y+z)]_q  = [ x\,y + x\,z]_q  \nonumber \\
& = & (x\,y)_q \oplus_q (x \,z)_q \nonumber \\
& = & (x_q \Diamond_q y_q) \oplus_q (x_q \Diamond_q z_q).
\end{eqnarray}


\section{$q$-Pascal's triangles}

\label{sec:pascaltriangle}

The deformations of operations and numbers open questions about
other mathematical objects derived from them. An interesting class
of those objects are the Pascal's triangles. Recently, some works
connect nonextensive statistical mechanics with Leibnitz \cite{Sato}
and Pascal's triangles derived from the $q$-product \cite{Suyari}.

In order to exhibit some simple applications of such deformations,
in this section we construct $q$-Pascal's triangles using $q$-sum as
the deformed operation. In this way, their elements are $q$-numbers.
Different patterns are illustrated for different values of $q$. For
example, we present $q$-Pascal's triangles for $q=0$, $q=1.5$, $q=2$
and $q=3$:

\begin{itemize}

\item [] a) {\bf Increasing pattern}

For $q=0$, we obtain:

\begin{figure}[ht!]
\begin{equation}
\begin{array}{ccccccccccccc}
& & & & & & 1   \\
& \\
& & & & & 1 & & 1   \\
& \\
& & & & 1 & & 3 & & 1  \\
& \\
 & & & 1 & & 7 & & 7 & & 1 \\
 & \\
& & 1 \,\,\, & & 15 & & 63 & & 15 & & \,\, \, 1 \\
& \\
& 1 \,\,\, \,& & 31 & & 1023 & & 1023 & & 31 & &\, \,\,\, 1 \\
& \\
 1 \, \,\,\, \,\,& & 63 \,\,\, & & 32767 & & 1048575 & & 32767 & & \,\,\, 63 & & \,\,\,\,\, \,1 \\
 \nonumber
\end{array}
\end{equation}
\end{figure}

Note that this increasing pattern occurs for any value of $q \le 1$.
For $q=1$, we recover the usual Pascal's triangle. It is compatible
with the divergent curve shown in figure \ref{fig1} for natural
values of $x$.

\vspace{0.5cm}

\item [] b) {\bf Asymptotical pattern}

For $q=1.5$, we obtain:

\begin{equation}
\begin{array}{ccccccccccccccccccccc}
  & & & & & & & & 1   \\
& \\
  & & & & & & & 1 & & 1   \\
& \\
  & & & & & & 1 & & 1.5 & & 1  \\
& \\
   & & & & & 1 & & 1.75 & & 1.75 & & 1 \\
 & \\
 & & & & 1 & & 1.875 & & 1.968 & & 1.875 & &  1 \\
& \\
  & & & 1  & & 1.937 & & 1.998 & & 1.998 & & 1.937 & &  1 \\
& \\
  & &  1  & & 1.968 & & 1.999 & & 1.999 & & 1.999 & & 1.968 & &  1 \\
 & \\
 & 1 & & 1.984 & & 1.999 & &  2 & &  2 & &  1.999 & & 1.984 & & 1 \\
& \\
  1\,\,\,\,\,\,\,\, & & 1.992 & &  2 & &  2 & &  2 & &  2 & &  2 & & 1.992 & & \,\,\,\,\,\,\,\, 1 \\
  \nonumber
\end{array}
\end{equation}

\vspace{0.5cm}

 \item [] c) {\bf Fixed pattern}

For $q=2$, we obtain:

\begin{equation}
\begin{array}{ccccccccccccc}
& & & & & & 1   \\
& & & & & 1 & & 1   \\
& & & & 1 & & 1 & & 1  \\
 & & & 1 & & 1 & & 1 & & 1 \\
& & 1 & & 1 & & 1 & & 1 & &  1 \\
& 1  & & 1 & & 1 & & 1 & & 1 & &  1 \\
 1  & & 1 & & 1 & & 1 & & 1 & & 1 & &  1 \\
 \nonumber
\end{array}
\end{equation}

For any value of $1<q<2$, the elements are positive greater than 1.
In the limit case $q=2$, it converges to the fixed pattern shown
above. In general, if $1<q<3$, $\lim_{n \rightarrow \infty} n_q =
1/(q-1)$; for $q=1.5$, $\lim_{n \rightarrow \infty} n_{1.5} = 2$;
for $q=2$, $n_2 =1$ for any value of n.

\vspace{0.5cm}

\item [] d) {\bf Self-similar pattern}

For $q=3$, we obtain:

\begin{equation}
\begin{array}{ccccccccccccccccccccccccccccc}
& & & & & & & & & & & & & & 1   \\
& & & & & & & & & & & & & 1 & & 1   \\
& & & & & & & & & & & & 1 & & {\bf 0} & & 1  \\
 & & & & & & & & & & &  1 & & 1 & & 1 & & 1 \\
& & & & & & & & & & 1 & & {\bf 0} & & {\bf 0} & & {\bf 0} & &  1 \\
& & & & & & & & & 1  & & 1 & & {\bf 0} & & {\bf 0} & & 1 & &  1 \\
 & & & & & & & & 1  & & {\bf 0} & & 1 & & {\bf 0} & & 1 & & {\bf 0} & &  1 \\
 & & & & & & & 1  & & 1 & & 1 & & 1 & & 1 & & 1 & &  1 & & 1 \\
 & & & & & & 1 & & {\bf 0} & & {\bf 0} & &  {\bf 0} & &  {\bf 0} & & {\bf 0} & & {\bf 0} & & {\bf 0} & & 1 \\
 & & & & & 1 & & 1 & &  {\bf 0} & & {\bf 0} & & {\bf 0} & & {\bf 0} & & {\bf 0} & & {\bf 0} & & 1 & & 1 \\
  & & & & 1 & & {\bf 0} & & 1 & & {\bf 0} & & {\bf 0} & & {\bf 0} & & {\bf 0} & & {\bf 0} & & 1 & &  {\bf 0} & & 1\\
 & & & 1 & &  1 & &  1 & &  1 & & {\bf  0} & &  {\bf 0} & & {\bf  0} & &  {\bf 0} & &  1 & &  1 & & 1 & & 1 \\
 & & 1 & & {\bf 0} & & {\bf 0} & & {\bf 0} & & 1 & & {\bf 0} & & {\bf 0} & & {\bf 0} & & 1 & & {\bf 0} & & {\bf 0} &
 & {\bf 0} & & 1 \\
 & 1 & & 1 & & {\bf 0} & & {\bf 0} & & 1 & & 1 & & {\bf 0} & & {\bf 0} & & 1 & & 1 & & {\bf 0} & &
{\bf 0} & & 1 & & 1 \\
 1 & & {\bf 0} & & 1 & & {\bf 0} & & 1 & & {\bf 0} & & 1 & & {\bf 0} & & 1 & &
 {\bf 0} & & 1 & & {\bf 0} & & 1 & & {\bf 0} & & 1  \nonumber
\end{array}
\end{equation}

For any value of $2< q <3$, the elements are positive numbers
smaller than 1. In the limit case  $q=3$, the $3$-Pascal's triangle
presents a self-similar pattern due to the isomorphism between
${\mathbb Z}_{q=3}$ and ${\mathbb Z} \,\, mod \,\, 2$ shown in last
section.

\end{itemize}


\section{Conclusions and Perspectives}

\label{sec:conclusions}

In this work, we explore the properties of the algebraic structure
derived from the $q$-sum which implies a new product, in a natural
way, that recovers the distributive property. It is done by
constructing the $q$-numerical sets based on $q$-sum. We show that,
assuming some properties, the $q$-product does not distribute over
any sum. Therefore, using the $q$-numbers, we define a new deformed
product, called $\Diamond_q$ that distributes over the $q$-sum.
Finally, different patterns of Pascal $q$-triangles, whose elements
are $q$-numbers, are shown.

Our results illustrate the diversity of mathematical structures that
may be derived from the deformation of operations and numbers. It is
interesting that the nonextensive statistical mechanics called the
attention to deformations that were studied in the context of
Mathematics as well as some known mathematical objects  as Heine
number and Pascal's triangles. This work is a motivation of
investigating more deeply the connections between nonextensive
statistical mechanics and mathematical structures.

\vspace{0.3cm}


\section*{Acknowledgments}

This work is partially supported by CNPq -- Conselho Nacional de
Desenvolvimento Cient\'{\i}fico e Tecnol\'ogico (Brazilian Agency).
We acknowledge  Ricardo Alc\^antara, Luciano Dias and Wagner Telles
for fruitful discussions.


\end{document}